\documentstyle[12pt]{article}
\begin{document}

\title{MEASURING MULTIPOLE MOMENTS OF WEYL METRICS BY MEANS OF GYROSCOPES }

\author{L. Herrera\thanks{Also at UCV, Caracas, Venezuela. E-mail address:
lherrera@gugu.usal.es}\\Area de F\'\i
sica Te\'orica. Facultad de Ciencias.\\ Universidad de Salamanca. 37008
Salamanca, Espa\~na. \\ and\and J. L. Hern\'andez
Pastora \\Area de F\'\i sica Te\'orica. Facultad de Ciencias.\\ Universidad de
Salamanca. 37008 Salamanca, Espa\~na.}

\date{}
\maketitle

\begin{abstract}
Using the technique of Rindler and Perlick  we calculate the total
precession per revolution of a gyroscope circumventing the source of Weyl
metrics. We establish thereby a link between the multipole moments of the
source and an ``observable'' quantity.
Special attention deserves the case of the $\gamma$-metric. As an extension
of this result we also present the corresponding expressions for some
stationary space-times.
\end{abstract}

\newpage
\section{Introduction}
Weyl exterior solutions to Einstein equations \cite{weyl}, represent all
possible static axially symmetric space-times in the
context of general relativity.
They may be represented as series expansions of suitable defined
relativistic multipole moments \cite{yo}. Therefore, any of the
Weyl metrics are, in principle, characterized by a specific combination of
such multipoles. One way to provide physical content
to those solutions consists in establishing a link between their multipole
moments and quantities measured from well defined and
physically reasonable experiments. It is the purpose of this work to
establish such a link, by calculating the total precession per
 revolution of gyroscopes circumventing the symmetry axis. By doing so, the
multipole moments of different Weyl metrics become
 ``measurable'' in the sense that they are expressed through quantities
obtained from well defined and physically reasonable
 experiments (we are of course not discussing about the actual technical
feasability of such experiments). These results illustrate
 further the usefulness of gyroscopes in the study of gravitational
phenomena \cite{luisyo},
 \cite{luisotros}.

All calculations are carried out using the method proposed by Rindler and
Perlick \cite{rinper}, a brief resume of which is given in the next
section, together with the notation and the specification of the space-time
under consideration.

In section $3$ we obtain, for a selection of Weyl metrics,  the precession
per revolution relative to the original frame of a gyroscope rotating round
the axis of symmetry. For sake of generality we present in Section $4$ some
results concerning stationary metrics.  Finally, the results are discussed
in the last section.

\section{The space-time and the Rindler-Perlick -
 method}

\subsection{The Weyl metrics}

Static axysymmetric solutions to Einstein equations are given by the Weyl
metric \cite{weyl}
\begin{equation}
ds^2 = -e^{2 \Psi} dt^2 + e^{-2 \Psi} [e^{2 \Gamma}(d \rho^2 +dz^2)+\rho^2
d \phi^2 ]
\label{elin}
\end{equation}
where metric functions have to satisfy
\begin{equation}
\Psi_{, \rho \rho}+\rho^{-1} \Psi_{, \rho}+\Psi_{, zz} = 0
\label{meq1}
\end{equation}
and
\begin{equation}
\Gamma_{, \rho}= \rho (\Psi_{, \rho}^2-\Psi_{, z}^2) \quad; \quad \Gamma_{,
z}= 2 \rho \Psi_{, \rho} \Psi_{, z}
\label{meq2}
\end{equation}

Observe that  (\ref{meq1}) is just the Laplace equation for $\Psi$ (in the
Euclidean space), and furthermore it represents the integrability condition
for (\ref{meq2}), implying that for any '' Newtonian'' potential we have a
specific Weyl metric, a well known result.

The general solution of the Laplace equation (\ref{meq1}) for the function
$\Psi$, presenting an asymptotically flat behaviour, results to be
\begin{equation}
\Psi = \sum_{n=0}^{\infty} \frac{a_n}{r^{n+1}} P_n(\cos \theta)
\label{psi}
\end{equation}
where $r=(\rho^2+z^2)^{1/2}$, $\cos \theta= z/r$ are Weyl spherical
coordinates and $P_n(\cos \theta)$ are Legendre Polynomyals. The
coefficients $a_n$ are arbitrary real constants which have been named in
the literature ``Weyl moments'', although they cannot be identified as
relativistic multipole moments in spite of the formal similarity between
expression (\ref{psi}) and the Newtonian potential. Then, equations
(\ref{meq2}) are solved to give function $\Gamma$ in terms of Weyl moments
as follow \cite{weyl}
\begin{equation}
\Gamma = \sum_{n,k=0}^{\infty} \frac{(n+1)(k+1)}{n+k+2} \frac{a_n
a_k}{r^{n+k+2}} (P_{n+1}P_{k+1}- P_n P_k)
\label{ga}
\end{equation}

Another interesting way of writting the solution (\ref{psi}) was obtained
by Erez-Rosen \cite{erroz} and Quevedo \cite{quev}, integrating equations
(\ref{meq1}, \ref{meq2}) in prolate spheroidal coordinates, which are
defined as follows
\begin{eqnarray}
x & = & \frac {r_{+}+r_{-}}{2 \sigma} \quad , \quad y  =  \frac
{r_{+}-r_{-}}{2 \sigma}\nonumber \\
r_{\pm} & \equiv & [\rho^2+(z\pm \sigma)^2]^{1/2} \nonumber \\
x & \geq & 1 \quad , \quad -1 \leq y \leq 1
\label{pro}
\end{eqnarray}
where $\sigma$ is an arbitrary constant which will be identified later with
the Schwarzschild's mass. Inverse relation between both families of
coordinates is given by
\begin{eqnarray}
\rho^2 & = & \sigma^2(x^2-1)(1-y^2) \nonumber \\
z & = & \sigma \, x \, y
\label{invpro}
\end{eqnarray}
The prolate coordinate $x$ represents a radial coordinate, whereas the
other coordinate, $y$ represents the cosine function of the polar angle.

In these prolate spheroidal coordinates the Weyl metric is given by
\begin{eqnarray}
ds^2 & = & -e^{2 \Psi} dt^2 \nonumber \\
& + & \sigma^2e^{-2 \Psi} \left[ e^{2
\Gamma}(x^2-y^2)(\frac{dx^2}{x^2-1}+\frac{dy^2}{1-y^2})+(x^2-1)(1-y^2)d\phi^
2 \right]
\label{prome}
\end{eqnarray}

Then, the corresponding equations that the metric functions $\Gamma$ and
$\Psi$ have to satisfy, can be solved to obtain for $\Psi$
\begin{equation}
\Psi = \sum_{n=0}^{\infty} (-1)^{n+1} q_n Q_n(x) P_n(y)
\label{propsi}
\end{equation}
being $Q_n(y)$ Legendre functions of second kind and $q_n$ a set of
arbitrary constants.

Both sets of coefficients, ${a_n}$ and ${q_n}$, characterize any Weyl
metric \cite{quev}. Nevertheless these constants do not give us physical
information about the metric since they do not represent the ``real''
multipole moments of the source. That is not the case for the relativistic
multipole moments defined by Geroch \cite{ger}, Hansen \cite{han} and
Thorne \cite{th}, which , as it is known, characterize completely and
uniquely, at least in the neighbourhood of infinity, every asymptotycally
flat and stationary vacuum solution  \cite{kun1} , \cite{kun2} providing at
the same time a physical description of the corresponding solution.

An algorithm to calculate the Geroch multipole moments was developed by
G.Fodor, C. Hoenselaers and Z. Perjes \cite{fhp} (FHP). By applying such
method, the resulting multipole moments of the solution are expressed in
terms of the Weyl moments. Similar results are obtained from the Thorne's
definition, using harmonic coordinates.  The structure of the obtained
relation between coefficients $a_n$ and these relativistic moments allows
to express the Weyl moments as a  combination of the Geroch relativistic
moments. For instance, the first coefficients result to be
\begin{eqnarray}
a_0 & = & -M_0  \\
a_2 & = & -\frac 13 M_0^3-M_2  \\
a_4 & = & -\frac 15 M_0^5-\frac 87 M_0^2 M_2
\label{an}
\end{eqnarray}
where only massive multipole moments ($M_n$) appear since the metrics are
considered to posses equatorial symmetry.
Obviously, the general relation between Weyl moments and relativistic
multipole moments are not known. Nevertheless, choosing some specific
multipole moments, it is possible to obtain the whole set of Weyl moments
needed to define a solution containing the required multipole structure.
Thus, for example, one of us \cite{yo} obtained the relativistic vacuum
solution corresponding to an object consisting exclusively of mass and
quadrupole moment. This metric will be treated later in our analysis.

\subsection{The Rindler-Perlick method}

This method consists in transforming the angular coordinate $\phi$ by
\begin{equation}
\phi = \phi^{\prime}+ \omega t,
\label{phi}
\end{equation}
where $\omega$ is a constant. Then the original frame is replaced by a
rotating frame. The transformed metric is written in a canonical form, (we
have slightly changed the original notation in \cite{rinper} to avoid
confussion with our notation)
\begin{equation}
\displaystyle{ds^2 = -e^{2 \Phi}(dt- \omega_i dx^i)^2+h_{ij} dx^idx^j},
\label{cano}
\end{equation}
with latin indexes running from $1$ to $3$ and $\Phi$, $\omega_i$ and
$h_{ij}$ depend on the spatial coordinate $x^i$ only (we are omitting
primes). Then, it may be shown that the four acceleration $A_{\mu}$ and the
rotation three vector $\Omega^i$ of the congruence of world lines
$x^i=$constant are given by \cite{rinper},
\begin{eqnarray}
A_{\mu} & = & (0, \Phi_{, i}) \\
\Omega^i & = & \frac 12 \displaystyle{e^{\Phi}}(det \,
h_{mn})^{-1/2}\epsilon^{ijk} \omega_{k , j},
\label{accyrot}
\end{eqnarray}
where the comma denotes partial derivative. It is clear from the above that
if $\Phi_{, i}=0$, then particles at rest in the rotating frame follow a
circular geodesic. On the other hand, since $\Omega^i$ describes the rate
of rotation with respect to the proper time at any point at rest in the
rotating frame, relative to the compass of inertia, then $-\Omega^i$
describes the rotation of the compass of inertia (the {\it gyroscope}) with
respect to the rotating frame. Applying  (\ref{phi}) to the original frame
of (\ref{elin}) written in Weyl's spherical coordinates, i.e,
\begin{equation}
ds^2 = -f dt^2+ f^{-1} [e^{2 \Gamma}(dr^2+r" d \theta^2)+r^2 \sin^2 \theta
d\phi^2]
\label{elin2}
\end{equation}
 we cast (\ref{elin2}) into the canonical form (\ref{cano}) where
\begin{eqnarray}
e^{2 \Phi} & = & f - \omega^2 f^{-1} r^2 \sin^2 \theta  \\
\omega_i & = & \displaystyle{e^{-2 \Phi}} \left( 0, 0, \omega f^{-1} r^2
\sin^2 \theta \right)  \\
h_{rr} & = & f^{-1} e^{2 \Gamma}  \\
h_{\theta \theta} & = & r^2 f^{-1} e^{2 \Gamma}  \\
h_{\phi \phi} & = &\displaystyle{ e^{-2 \Phi}} r^2 \sin^2 \theta
\label{canele}
\end{eqnarray}
whit $f \equiv e^{2 \Psi}$.
If we take into account the condition for circular geodesics, it results
\begin{equation}
\omega^2 = {\displaystyle \frac{f^2 (f_{, r}+f_{, \theta})}{2 r f \sin
\theta\left(\sin \theta+r \cos \theta \right)-r^2 \sin^2 \theta \left
(f_{, r}+f_{, \theta} \right)}}
\label{wgeod}
\end{equation}

If we consider circular geodesics in the equatorial plane, then the
parameter $\omega$ turns out to be
\begin{equation}
\omega = {\displaystyle \frac{f_{, r}^{1/2} f}{\left( 2 r f - r^2 f_{, r}
\right)^{1/2}}}
\label{wgeeq}
\end{equation}

\subsection{Rate of precession}

The expression for the rate of precession $\Omega \equiv \left( \Omega^i
\Omega^j h_{ij}\right)^{1/2}$ results to be
\begin{equation}
\Omega = \displaystyle{\frac{f^{1/2} e^{- \Gamma} \omega \left [\left(f_{,
\theta} \sin \theta- f  \cos \theta \right)^2+\sin^2 \theta \left(r f_{,
r}-f \right)^2 \right]^{1/2}}{f^2-\omega^2 r^2 \sin^2 \theta}}
\label{omega}
\end{equation}

Taking into account only circular geodesics in the equatorial plane, then
we get from (\ref{wgeeq}) and (\ref{omega})
\begin{equation}
\Omega = \displaystyle{\frac 12 \left(\frac {f_{, r}}{f}\right)^{1/2}
\frac{e^{- \Gamma}}{r^{1/2}} \left(2 f -r f_{, r} \right)^{1/2}}
\label{omegaeq}
\end{equation}

According to the meaning of $\Omega$ given above, it is clear that the
orientation of the gyroscope, moving around the axis of symmetry, after one
revolution, changes by
\begin{equation}
\triangle \phi^{\prime} = - \Omega \triangle \tau
\label{fase}
\end{equation}
where $\triangle \tau$ is the proper time interval corresponding to one
period. Then from the canonical form of the metric (\ref{cano}),
\begin{equation}
\triangle \phi^{\prime} = -2 \pi \frac{ \Omega e^{\Phi}}{\omega},
\label{fase2}
\end{equation}
as measured in the rotating frame. In the original system we have
\begin{equation}
\triangle \phi =2 \pi \left( 1-\frac{ \Omega e^{\Phi}}{\omega} \right).
\label{fase}
\end{equation}

This quantity $\triangle \phi$, calculated over one revolution around a
circular orbit for the Weyl metrics,  results to be
\begin{equation}
\triangle \phi = 2 \pi \left( 1- \frac{e^{-\Gamma} \omega \left[\left( f_{,
\theta} \sin \theta- f  \cos \theta \right)^2+\sin^2 \theta \left( r f_{,
r}-f \right)^2 \right]^{1/2}}{\left[ f^2-\omega^2 r^2 \sin^2 \theta
\right]^{1/2}} \right).
\label{faseweyl}
\end{equation}

And considering a circular geodesic in the equatorial plane we have,
\begin{equation}
\triangle \phi =2 \pi \left( 1- \frac{{\sqrt 2}}{2}e^{-\Gamma}f^{-1}
\left[(f-r f_{, r})(2f -r f_{, r})\right]^{1/2}  \right).
\label{faseweq}
\end{equation}

In terms of the metric function $\Psi$ we have,
\begin{eqnarray}
\Omega & = & \frac{1}{r^{1/2}} e^{-\Gamma} e^{\Psi} \Psi_{, r}^{1/2} (1- r
\Psi_{, r})^{1/2} \\
\triangle \phi & = & 2 \pi \left( 1- e^{-\Gamma} (1-r \Psi_{, r})^{1/2}
(1-2 r \Psi_{, r})^{1/2} \right).
\label{psitr}
\end{eqnarray}

Since the metric functions $\Psi$ and $\Gamma$ are known for any Weyl
solution, we can obtain both the rate of precession $\Omega$ and $\triangle
\phi$, using (\ref{psi}) and (\ref{ga}), in terms of the Weyl moments of
the solution. Up to order ${\cal O}(r^{-5})$ we have the following result
\begin{eqnarray}
\Omega & = &1+\frac 52 a_0 \frac 1r + \frac {19}{8} a_0^2 \frac 1{r^2}+
\left( \frac {107}{48} a_0^3 - \frac{11}{4} a_2 \right) \frac 1{r^3}
\nonumber \\
& + & \left( \frac {457}{384} a_0^4 - \frac{31}{8} a_2 a_0-\frac 14
a_1^2\right) \frac 1{r^4}+{\cal O}(r^{-5})
\label{omytri0}
\end{eqnarray}

\vspace*{3mm}

\begin{eqnarray}
\triangle \phi & = & 2 \pi \left[- \frac 32 a_0 \frac 1r -\frac 38 a_0^2
\frac 1{r^2}-\left( \frac{15}{16} a_0^3- \frac 94 a_2 \right) \frac 1{r^3}
\right.
\nonumber \\
& -& \left.  \left( \frac{3}{8} a_0 a_2+\frac{29}{128} a_0^4 \right) \frac
1{r^4}+{\cal O}(r^{-5})\right].
\label{omytri}
\end{eqnarray}

Now, making use of the fact that the Weyl moments are some combination of
the multipole moments (\ref{an}) we can obtain the above quantities in term
of the relativistic massive moments $M_n$ as follows,
\begin{eqnarray}
\Omega & = &1-\frac 52 M_0 \frac 1r + \frac {19}{8} M_0^2 \frac 1{r^2}+
\left( \frac{11}{4} M_2 -\frac {151}{48} M_0^3 \right) \frac 1{r^3}
\nonumber \\
& - & \left( \frac {3}{128} M_0^4 + \frac{31}{8} M_2 M_0+\frac 14
M_1^2\right) \frac 1{r^4}+{\cal O}(r^{-5})
\label{psitrmom20}
\end{eqnarray}

\vspace*{3mm}

\begin{eqnarray}
\triangle \phi & = & 2 \pi \left[ \frac 32 M_0 \frac 1r -\frac 38 M_0^2
\frac 1{r^2}-\left(\frac 94 M_2-\frac{3}{16} M_0^3 \right) \frac 1{r^3}
\right. \nonumber \\
& + & \left. \left( \frac{3}{8} M_0 M_2+\frac{45}{128} M_0^4 \right) \frac
1{r^4} \right].
\label{psitrmom2}
\end{eqnarray}

In the next section we shall specialize eqs. (\ref{psitrmom20}) and
(\ref{psitrmom2}) to some specific solutions.

\section{Some examples of Weyl solutions}

\subsection{ Curzon metric}

This solution of the Weyl's family \cite{curzon} corresponds to a function
$\Psi$ with only the first Weyl moment. Taking all the coefficients $a_n$
equal to zero, except $a_0$, then we have the $\triangle \phi$ in terms of
the unique parameter of this metric
\begin{equation}
\triangle \phi = 2 \pi \left[ - \frac 32 a_0 \frac 1r -\frac 38 a_0^2 \frac
1{r^2}- \frac{15}{16} a_0^3 \frac 1{r^3}-\left( \frac{3}{8} a_0
a_2+\frac{29}{128} a_0^4 \right) \frac 1{r^4} \right].
\label{tricu}
\end{equation}
This expression is exactly the expansion in power series of the inverse
radial coordinate of the quantity $\triangle \phi$, for this solution, with
metric functions $\Psi=\displaystyle{\frac{a_0}{r}}$ and
$\Gamma=-\displaystyle{\frac{a_0^2}{2 r^2}}$
\begin{equation}
\triangle \phi = 2 \pi \left[ 1- \frac 1r  \displaystyle{e^{\frac{a_0^2}{2
r^2}}} \sqrt{(r+a_0)(r+2 a_0)} \right]
\label{curtot}
\end{equation}

\subsection{Erez-Rosen solution}

In prolate spheroidal coordinates, this metric is given by the metric
function $\Psi$ of the form
\begin{equation}
\Psi = -q_0 Q_0(x)P_0(y)+q_2 Q_2(x)P_2(y),
\label{erezro}
\end{equation}
with $ q_0=1$.  The first term corresponds to the Schwarzschild metric, and
so, this metric posseses two parameters which represent the mass and the
quadrupole moment.
The relation between Weyl moments and the coefficients $q_n$ of the
function $\Psi$ is known \cite{yo2}, and therefore it is possible to use
expression (\ref{omytri}) to obtain an expansion of $\triangle \phi$.
Another way to proceed is to use the expression (\ref{psitrmom2}) with the
relativistic multipole moments involved, knowing that the mass and
quadrupole moment of this metric are respectively
\begin{eqnarray}
M_0 & = & \sigma  \\
M_2 & = & \frac 2{15} \sigma^3 q_2
\label{momerz}
\end{eqnarray}
where $\sigma$ is the Schwarzschild's mass. The result is the following
\begin{eqnarray}
\triangle \phi & = & 2 \pi \left[-\frac 32 \lambda+\frac 38 \lambda^2+
\left(\frac{3}{10} q_2-\frac{3}{16} \right) \lambda^3- \left(\frac{1}{20}
q_2+\frac{45}{128} \right) \lambda^4 \right. \nonumber \\
& + & \left.  \left(\frac{51}{560} q_2-\frac{69}{256} \right)
\lambda^5+{\cal O}(\lambda^6)\right]
\label{eso}
\end{eqnarray}
where $\lambda \equiv \displaystyle{\frac{\sigma}{r}}$.

$\triangle \phi$ for a circular geodesic orbit in the equatorial plane
expressed in  prolate spheroidal coordinates results to be
\begin{equation}
\triangle \phi  =  2 \pi \left[ 1-e^{-\Gamma}\left( \frac{ D_x(\Psi)-x}{2
D_x(\Psi)-x}\right)^{1/2} \left( 1+4 \frac { D_x(\Psi)}{x^2} (
D_x(\Psi)-x)\right)^{1/2}\right]
\label{tripro}
\end{equation}
with $ D_x(\Psi)  \equiv  (x^2-1) \Psi_{, x}$.

So, using (\ref{erezro}) and the corresponding function $\Gamma$, it turns
out to be
\begin{equation}
\triangle \phi = 2 \pi \left( 1-e^{-\Gamma}\frac{\sqrt{2}}{4 x} A^{1/2}
(A-2 x)^{1/2}\right)
\label{triprotot}
\end{equation}
where
\begin{equation}
A \equiv 4-3 q_2 x^3 \ln\left(\frac{x-1}{x+1}\right)-6 q_2 x^2-2 x+3 q_2 x
\ln \left(\frac{x-1}{x+1}\right)+4 q_2
\label{A}
\end{equation}
If we take $q_2=0$ then the Schwarzschild expression is obtained for
$\triangle \phi$, i.e.,
\begin{equation}
\triangle \phi = 2 \pi \left[1-\sqrt{\frac{x-2}{x+1}}\right]
\label{schw}
\end{equation}
or using the radial Schwarzschild coordinate $\hat r = \sigma (x+1)=\sigma
+\sqrt{r^2+ \sigma^2}$, it takes the well known form
\begin{equation}
\triangle \phi = 2 \pi \left[1-\sqrt{1-\frac{3 \sigma}{\hat r}}\right]
\label{schwrs}
\end{equation}

The expression (\ref{triprotot}) can be written in Weyl coordinates as a
power series in radial coordinate. Doing so one obtains the same result
that putting the multipole moments (\ref{momerz}) into the general
expression (\ref{psitrmom2}) for $\triangle \phi$.

Another useful expression, for the discussion below, results if one expands
(\ref{triprotot}) in power series of the parameter $q_2$.
 The order zero of the expansion corresponds to the Schwarzschild term
\begin{equation}
\triangle \phi = 2 \pi \left[ 1-\sqrt{\left(
\sqrt{1+\lambda^2}-\lambda\right) \left( \sqrt{1+\lambda^2}-2 \lambda
\right)}\right].
\label{schwrw}
\end{equation}
and the order $q_2$ gives the next contribution to  $\triangle \phi$, which
reads
\begin{eqnarray}
& \displaystyle{\frac 18 \frac 1{\sqrt{(x^2-1) (1-x) (2-x)}}} \left[ \ln
\left(\displaystyle{\frac{x-1}{x+1}}\right) (9 x^4+18 x^3+36 x-45 x^2)
\right. \nonumber \\
& +  \left. 64-84 x+\ln \left(\displaystyle{\frac{x^2-1}{x^2}}\right) (24
x-8 x^2-16) \right]
\label{orderq}
\end{eqnarray}

\subsection{Monopole-Quadrupole solution}

This is an exact solution of the static and axysymmetric Einstein vacuum
equations which is written as a series in a parameter $q$ representing the
dimensionless quadrupolar moment of the solution. The first term in the
expansion (order zero in the parameter $q$) corresponds to Schwarzschild,
whereas the whole series describes a solution which only posseses mass and
quadrupole moment \cite{yo}. The expression for the metric function $\Psi$
is the following
\begin{eqnarray}
\Psi & = & \sum_{\alpha=0}^{\infty} q^{\alpha} \Psi_{q^{\alpha}} \\
\Psi_{q^{\alpha}} & = & -\sum_{k=0}^{\alpha-1} b_k(\alpha)
\left[\frac{P_k^+}{(x+y)^{k+1}}+(-1)^k \frac{P_k^-}{(x-y)^{k+1}}\right]
\nonumber\\
& - & \sum_{k=0}^{\alpha}q_k(\alpha) Q_{2k}(x) P_{2k}(y)
\label{psiyo}
\end{eqnarray}
where the $b_k(\alpha)$ and $q_k(\alpha)$ are well defined coeffcients
\cite{yo}, and $P^{\pm}$ are Legendre polynomials with arguments
$P_n\left(\displaystyle{\frac{xy \pm 1}{x \pm y}}\right)$

Considering the first two terms in the expansion, one obtains a new exact
solution with two parameters representing the mass and quadrupole moment.
The extent to which this metric describes the field of a compact body very
close to an spherical mass is discussed in \cite{yo}. Here, we want to
calculate $\triangle \phi$ corresponding to this metric and to compare the
result with the obtained for the Erez-Rosen metric.

The parameter $q$ gives directly the value of the quadrupole moment and,
therefore, we obtain $\triangle \phi$ from (\ref{psitrmom2}) with the
values $M_0 = \sigma$ and $M_2 = q \sigma^{3}$. The expression for
$\triangle \phi$ in prolate coordinates results to be
\begin{equation}
\triangle \phi = 2 \pi \left( 1-e^{-\Gamma}\frac{\sqrt{2}}{24 x^5} B^{1/2}
(B-12 x^5)^{1/2} \right)
\label{triprotyo}
\end{equation}
where
\begin{equation}
B \equiv A x^4 + 2 x^2 (10 x^2-5 x^3-15 q)
\label{B}
\end{equation}
and  $q_2=\displaystyle{\frac{15}{2} q}$.

 Expanding in
 power series of $q$, in order to compare with Erez-Rosen, we obtain again for
 the order zero of the expansion the Schwarzschild term,
and the order $q$ gives the next contribution to the $\triangle \phi$, which is
\begin{eqnarray}
& \displaystyle{\frac 5{32} \frac 1{\sqrt{(x-2) (x+1)}x^2 (x-1)}} \left[
\ln \left(\displaystyle{\frac{x-1}{x+1}}\right) (78 x^5+9x^6+192 x^3-279
x^4) \right. \nonumber \\
& +  \left. 18 x^5+156 x^4+436 x^2-552x^3+104-168 x \right]
\label{orderqyo}
\end{eqnarray}

\subsection{$\gamma$-metric or Zipoy-Vorhees metric}

This metric deserves a special attention, since the singular structure of
its ``Newtonian'' potential $\Psi$, is the same as that of the
Schwarzschild solution (a line segment).
The metric functions defining this solution of Weyl's family are the
following \cite{zipo}, \cite{espo}, \cite{virb}
\begin{eqnarray}
\Psi & = & \frac{\gamma}{2} \ln \left(\frac{x-1}{x+1}\right)  \\
\Gamma & = & \frac{\gamma^2}{2} \ln \left(\frac{x^2-1}{x^2-y^2}\right)
\label{metgam}
\end{eqnarray}
As it is known the value $\gamma = 1$ yields the Schwarzschild solution.
The two first massive multipole moments of this metric, mass and quadrupole
moment,  are
\begin{eqnarray}
M_0 & = & \gamma \sigma  \\
M_2 & = & \gamma (1-\gamma^2) \frac{\sigma^3}{3}
\label{momgam}
\end{eqnarray}
From (\ref{momgam}) it follows that $\gamma > 1$ ($\gamma < 1$) correspond
to oblate (prolate) sources.
Putting these values into expression (\ref{psitrmom2}), we obtain
\begin{equation}
\triangle \phi = 2 \pi \left[\frac 32 \gamma \lambda -\frac38 \gamma^2
\lambda^2-\frac 34 \gamma (1-\frac54 \gamma^2) \lambda^3+ \frac 18 \gamma^2
(1+\frac{29}{16} \gamma^2) \lambda^4 \right]
\label{triga}
\end{equation}

Equivalently, this last result for $\triangle \phi$ can be obtained by
expanding  in power series of $\lambda$ the expression derived from
(\ref{tripro}) using  (\ref{metgam}), which is
\begin{equation}
\triangle \phi = 2 \pi \left(1- x^{\gamma^2-1} \sqrt{\frac{(x-\gamma)(x-2
\gamma)}{(x^2-1)^{\gamma^2}}}\right)
\label{trgapro}
\end{equation}
or in terms of $\lambda$
\begin{equation}
\triangle \phi = 2 \pi \left[1- (\sqrt{1+\lambda^2})^{\gamma^2-1}
\sqrt{(\sqrt{1+\lambda^2}-\gamma \lambda)(\sqrt{1+\lambda^2}-2 \gamma
\lambda)}\right]
\label{trgaweyl}
\end{equation}
this expression can be compared with the more familiar one written in terms
of the Schwarzschild radial coordinate $\hat r$
\begin{equation}
\triangle \phi = 2 \pi \left[1- (1-\Lambda)^{\gamma^2-1}
\sqrt{\frac{(1-\Lambda-\gamma \Lambda)(1-\Lambda-2\gamma
\Lambda)}{(1-2\Lambda)^{\gamma^2}}}\right]
\label{trgaweyl}
\end{equation}
where $\Lambda \equiv \displaystyle{\frac{\sigma}{\hat r}}$.

\subsection{Two stationary examples}

 An approximate solution of stationary Einstein vacuum field equations was
obtained in \cite{yo3} by means of an expansion of the
 Ernst potential in power series of the parameter ${\cal J} \equiv
\displaystyle{ i \frac{J}{M^2}}$, where $J$ is the total angular momentum
and $M$ is the mass. This metric with two parameters was constructed in
order to obtain a solution which posseses, up to the considered order in
${\cal J}$, mass and angular momentum in its multipolar structure.

Let us now  calculate $\triangle \phi$ for this metric and contrast the
result with the expression obtained from the Kerr metric. The metric
functions of the former (hereafter referred to as MJ) are
\begin{eqnarray}
f&=&\frac{x-1}{x+1} \nonumber \\
&+&  \frac{2 x j^2}{(x+1)^2} \left[ \frac{43 x^2-15 x^4-12}{16 x^3}
-\frac{15}{32} \frac{(x^4-1)}{x^2}  \ln \left(\frac{x-1}{x+1}\right)
\right]  \\
W&=&- 2  \frac{y}{x (x+1)^2} j
\label{mj}
\end{eqnarray}
where $E \equiv f+i W$ is the Ernst potential and $j \equiv
\displaystyle{\frac{J}{M^2}}$.  The other metric functions, in the
equatorial plane, are
 \begin{eqnarray}
\omega_{\phi} & = & 2 \sigma \left[\left( \frac 1{x+1}\right)+ \ln
\left(\frac{x-1}{x} \right) j \right] +{\cal O}(j^3)  \\
\Gamma & = & \frac 12 \ln \left(\frac{x^2-1}{x^2}\right)+{\cal O}(j^2)
\label{mjf}
\end{eqnarray}

As can be seen from above, the first correction to Schwarzschild appears at
order $j$ in $W$, and so, since we want to compare with Kerr, we will
calculate $\triangle \phi$ up to order $j$.

Since in both, Kerr  and MJ, the original lattice rotates with respect to a
compass of inertia, then a gyroscope fixed at radius $R$ in the original
lattice precesses with respect to neighboring points, in each case, at a
rate given by
 \begin{eqnarray}
\Omega_{Kerr} & = & \displaystyle{\frac{a \sigma}{R^3
\left(1-\displaystyle{\frac{2 \sigma}{R}}\right)}} \\
\Omega_{MJ} & = & \frac{j}{\sigma x (x-1) (x+1)^2}
\label{vortis}
\end{eqnarray}
being $R$ the Boyer-Lindquist radial coordinate.
Then, within  proper time $\triangle \tau = \sqrt{-g_{00}} \triangle t$ the
original lattice changes its orientation with respect to neighboring points
by
 \begin{eqnarray}
\hat{\triangle \phi}_{Kerr} & = &- \displaystyle{\frac{a \sigma}{R^3
\left(\sqrt{1-\frac{2 \sigma}{R}}\right)}}\triangle t \label{desfas0}\\
\hat{\triangle \phi}_{MJ} & = &- \frac{ j}{\sigma x (x-1)^{1/2}
(x+1)^{3/2}} \triangle t
\label{desfas}
\end{eqnarray}

Considering now a gyroscope orbiting in a circular geodesic, then from the
condition $\Phi_{,R}=0$ we obtain for the angular velocity
 \begin{eqnarray}
\omega_{Kerr} & = & \left(a+\sqrt{\frac{R^3}{\sigma}}\right)^{-1}  \\
\omega_{MJ} & = & \frac1{\sigma (x+1)^{3/2}}+ j \frac 1{x(x+1)^3}\left[-1+2
x \ln \left(\frac{x}{x-1}\right)\right]
\label{ws}
\end{eqnarray}

Making use of the Rindler-Perlick method one can easily obtain the total
precession per revolution $\triangle \phi$, with respect to the original
lattice,  of a gyroscope carried along a circular geodesic in the
equatorial plane. For MJ the result is
\begin{equation}
\triangle \phi = 2 \pi
\left[1-\displaystyle{\sqrt{\frac{x-2}{x+1}}}+\frac{1}{x(x+1)(x-2)^{1/2}} j
\right]
\label{desfmj}
\end{equation}
whereas for Kerr, we recover the well known expression
\begin{equation}
\triangle \phi = 2 \pi \left[1-\displaystyle{\sqrt{1-\frac{3 \sigma}{R}+2 a
\sqrt{\frac{\sigma}{R^3}}}} \right]
\label{desfker}
\end{equation}

As we are calculating $\triangle \phi$ for MJ solution up to order $j$ we
will handle last expression up to order $a$ (which is equivalent to
consider small values of $\displaystyle{\frac aR}$), and so we obtain the
Schiff precession term
\begin{equation}
\triangle \phi = 2 \pi \left[1-\sqrt{1-\frac{3 \sigma}{R}}- \frac aR
\sqrt{\frac{\sigma}{R}} \left(1-\frac{3 \sigma}{R} \right)^{-1/2}  \right]
\label{desfkeroa}
\end{equation}
And the expression (\ref{desfmj}) for MJ turns out to be in Boyer-Lindquist
coordinates
\begin{equation}
\triangle \phi = 2 \pi \left[1-\sqrt{1-\frac{3 \sigma}{R}}+\frac{\sigma}{R}
\frac aR \displaystyle{\sqrt{\frac{\sigma}{R}}}
\frac{1}{\left(1-\displaystyle{\frac{\sigma}{R}}\right)\left(1-\displaystyle
{\frac{3 \sigma}{R}}\right)^{1/2}}  \right]
\label{desfmjoa}
\end{equation}
(where the definition of $j$ and the fact that $a$ is the the angular
momentum per mass unit ,  have been used).
As can be seen from expressions (\ref{desfkeroa}) and (\ref{desfmjoa}) the
contribution to $\triangle \phi$ from the angular momentum for MJ is less
than  the Schiff precession term for Kerr, since
$\displaystyle{\frac{\sigma}{R}}$ is eventually small.

If that is so, we can expand both expressions in power series of
$\displaystyle{\frac{\sigma}{R}}$ and keeping only the order ${\cal
O}\left(\displaystyle{\frac{\sigma}{R}}\right)^{3/2}$, to obtain

\begin{equation}
\triangle \phi \approx \frac{3 \pi \sigma}{R} -2 \pi \frac aR
\sqrt{\frac{\sigma}{R}} -3 \pi \frac{a \sigma}{R^2}\sqrt{\frac{\sigma}{R}}
\label{schif}
\end{equation}
for the Kerr metric. And for MJ,
\begin{equation}
\triangle \phi \approx \frac{3 \pi \sigma}{R} +2 \pi \frac{a \sigma}{R^2}
\sqrt{\frac{\sigma}{R}}
\label{schifmj}
\end{equation}

Observe that $\triangle \phi$ as given by  (\ref{desfker}) and
(\ref{desfmj}) represents the precession of the gyroscope with respect to
the neighboring points of the original lattice and not with respect to a
compass of inertia. If the precession of the orbiting gyroscope is wanted
with respect to a fixed gyroscope, then  (\ref{desfker}) and (\ref{desfmj})
have to be reduced by the values (\ref{desfas}, \ref{desfas0})
respectively.

For a coordinate time $\triangle t = 2 \pi/ \omega$, the $\hat{\triangle
\phi}$ of the original lattice itself results from (\ref{desfas}) and
(\ref{ws}), to be
 \begin{eqnarray}
\hat{\triangle \phi}_{Kerr} & = & - 2 \pi \frac aR \sqrt{\frac{\sigma}{R}}
\left(1+\frac{\sigma}{R} \right)   \\
\hat{\triangle \phi}_{MJ} & = &  - 2 \pi \frac aR \sqrt{\frac{\sigma}{R}}
\left(1+2\frac{\sigma}{R} \right)
\label{wsordera}
\end{eqnarray}

As can be seen, the order of magnitud of these quantities is the same, and
so, one observes from  (\ref{schif}) and (\ref{schifmj}) that, whether one
compares the precession with respect to a fixed point or to neighboring
points of the original lattice, $\triangle \phi$ is larger for Kerr than
for MJ solution. As it follows from (\ref{schif}) and (\ref{schifmj}) the
ratio of nonspherical contributions to $\triangle \phi$, of both metrics,
is of the order $\displaystyle{\frac{\sigma}{R}}$. More important is the
fact that they have different signs.

\section{Conclusions}

We have established a relationship between the total precession per
revolution of a gyroscope circumventing the source and the relativistic
multipole moments describing the space-time. This result allows for
``measuring'' (in principle) the multipole moments (at least the first
ones) of a given source. Indeed, by displaying an array of gyroscopes along
the radial coordinate, circumventing the source, we obtain the curve
$\triangle  \phi = \triangle \phi (r)$, which leads to the values of the
coefficients in (\ref{psitrmom2}) by adjusting parameters.

Secondly, our results open the possibility to compare different
axysymmetric solutions in terms of an ``observable'' quantity ($\triangle
\phi$), and thereby to decide what space-time is actually ``in place''.

Thus for example, considering a neutron star (N-S) as a non-rotating source
of the Erez-Rosen metric, we can use expressions  (\ref{orderq}) and
(\ref{orderqyo}) to evaluate  the first contribution (and dominant, since
$q \sim 1.8 \times 10^{-4}$, assuming for the N-S the same eccentricity of
the sun ) to the $\triangle \phi$ due to the quadrupole moment.  Typical
values for a N-S are a mass like the sun and a radius of $10^{4} m$. So, we
obtain for the contribution of the quadrupole moment  $\triangle \phi \sim
9 \times 10^{-3} q \times 2 \pi$ for Erez-Rosen, and  $\triangle \phi \sim
- 7 \times 10^{-3} q \times 2 \pi$ for MQ solution, i.e, they are of the
same order but different sign.

In the case of the earth, the large value of $q$  ($\sim -3 \times
10^{15}$) renders the expansion in power series of $q$ useless. Instead of
that we may expand in power series of $\lambda$ which is now very small. We
obtain in this case that the contribution to  $\triangle \phi$ of the
quadrupole moment is  $ \sim - 2.5 \times 10^{-12}$ for both metrics.
If we consider  a N-S as a non-rotating source of the $\gamma$-metric,
then, with the values given above and being $\gamma = 0.9997$, the
$\triangle \phi$ is $\sim \displaystyle{\frac{\pi}{2}}$. The different
behaviour of Weyl  metrics with  respect to the $\triangle \phi$, is also
applicable to stationary metrics. In the two examples examined above we see
how differently, both contributions from the rotation of the source, are.

The relevance of the conclusions above becomes intelligible if we recall
the absence of a Birkhoff-type theorem for axysymetric solutions to
Einstein equations.

Finally we would like to make some additional comments on the
$\gamma$-metric. Because of (\ref{momgam}) it results that
\begin{equation}
\gamma =\frac{1}{\displaystyle{\sqrt{1+3 q}}}
\label{constrain}
\end{equation}
and therefore any possible source for the $\gamma$-metric should satisfy
the constraint $q > -1/3$. Thus for example, neither the earth nor the sun
(considered as non-rotating axysymmetric bodies) could serve as sources of
the $\gamma$-metric, since the corresponding values of $q$ are,
approximately, $-3 \times 10^{15}$ for the Earth and $-9 \times 10^{5}$ for
the Sun \cite{sol}.

\end{document}